\documentclass{desyproc}
\def\lappeq{\mathrel{\rlap{\raise.5ex\hbox{$<$}} {\lower.5ex\hbox{$\sim$}}}}

\begin{document}

\title{Particle Physics in the LHC Era and beyond}

\author{{\slshape Guido Altarelli}\\[1ex]
Dipartimento di Fisica `E.~Amaldi', Universit\`a di Roma Tre
\\
INFN, Sezione di Roma Tre, I-00146 Rome, Italy
\\
and
\\
CERN, Department of Physics, Theory Unit
\\
CH-1211 Geneva 23, Switzerland }

\contribID{47}

\confID{800}  
\desyproc{DESY-PROC-2009-xx}
\acronym{LP09} 
\doi  

\maketitle

\begin{abstract}
I present a concise review of where we stand in particle physics today. First, I will discuss QCD, then the electroweak sector and finally the motivations and the avenues for new physics beyond the Standard Model.
\end{abstract}

\section{Introduction}

This concluding talk is not meant to be a summary of the Symposium. Rather it is a very concise overview (as implied by the severe page limit) of the status of particle physics at the start of the LHC time \cite{HBurkhardt}, as reflected at this Conference, together with a collection of personal thoughts stimulated by the excellent talks that I followed in their totality.

In a few words the general map of particle physics is as follows. The Standard Model (SM) is a low energy effective theory (nobody can believe it is the ultimate theory). It happens to be renormalizable, hence highly predictive and is extremely well supported by the data. However, one expects corrections from higher energies, in particular already from the TeV scale (LHC!), and also from the GUT/Planck scales and possibly from some additional intermediate scales. But even as a low energy effective theory the SM is not satisfactory. In fact while QCD and the gauge part of the EW theory are well established, the Higgs sector is so far just a conjecture. Not only it needs an experimental verification but it introduces serious theoretical problems, like the hierarchy problem, that demand some form of new physics at the electroweak scale. The most important goals of the experiments at the LHC \cite{KJon-And, RCousins} are the clarification of the electroweak symmetry breaking mechanism, the search for signals of new physics at the TeV scale and, possibly, the identification of the unknown particles that make the dark matter in the Universe. 

The future of particle physics very much depends on the outcome of the LHC.
The LHC with the luminosity upgrade \cite{MNessi} will last for 15-20 years. Still the LHC cannot be all. A worldwide effort in neutrino physics is under way (T2K, DChooz, RENO, Daya Bay, NO$\nu$A......) \cite{SKopp}, \cite{YSuzuki},  \cite{JValle}. The continuation of experiments on the CKM mixing and CP violation \cite{SPrell}, \cite{TIijima} will take place at CERN with LHCb \cite{AGolutvin} and NA62..... \cite{TKomatsubara} and at new improved B-factories \cite{MGiorgi}. "Small" experiments of capital importance will produce their results like those on $\tau$ and charm decays \cite{YWang}, neutrino mass (e.g. KATRIN) and  neutrinoless double beta decay  \cite{CWeinheimer}, EDM's \cite{TMori} and the laboratory experiments on dark matter search \cite{NSmith}. A special mention deserves MEG, the on going search for the $\mu \rightarrow e \gamma$ decay at PSI \cite{TMori}, with a goal of improving the present bound by one or two orders of magnitude. They are now at the level of sensitivity of the present bound and will soon release the results from the ongoing run. We look forward to seeing the results because a positive signal would be a great discovery and is predicted in plausible models, like in some supersymmetric extensions of the SM. Astroparticle experiments, like Fermi-LAT,  PAMELA..... or AUGER, ICECUBE, ANTARES...., are more and more interesting for particle physics \cite{MMostafa}, \cite{MPunch}, \cite{PHulth}, also including the search for gravitational waves (VIRGO, LIGO....) \cite{FFidecaro} and experimental test of gravitation \cite{SHoedl}. For planning the next big step (ILC, CLIC...) \cite{EEisen} we must wait for the LHC outcome in  the first few years.


\section{QCD}

QCD stands as a main building block of the SM of particle physics. There are no essential problems of principle in its foundations and the comparison with experiment is excellent. 
For many years the relativistic quantum field theory of reference was QED, but at present QCD offers a more complex and intriguing theoretical laboratory. Indeed, due to asymptotic freedom, QCD can be considered as a better defined theory than QED. The statement that QCD is an unbroken renormalizable gauge theory, based on the $SU(3)$ colour group, with six kinds of triplets quarks with given masses, completely specifies the form of the Lagrangian in terms of quark and gluon fields. From the compact form of its Lagrangian one might be led to  think that QCD is a "simple" theory. But actually this simple theory has an extremely rich dynamical content, including the striking properties of asymptotic freedom and of confinement,  the complexity of the observed hadronic spectrum (with light and heavy quarks), the spontaneous breaking of (approximate) chiral symmetry, a complicated phase transition structure (deconfinement, chiral symmetry restoration, colour superconductivity), a highly non trivial vacuum topology (instantons, $U(1)_A$ symmetry breaking, strong CP violation,....), and so on.

So QCD is a complex theory and it is difficult to make its content explicit. Different routes have been developed over the years. There are non perturbative methods: lattice simulations (in great continuous progress), effective lagrangians valid in restricted specified domains, like chiral lagrangians, heavy quark effective theories, Soft Collinear Effective Theories (SCET), Non Relativistic QCD....) and also QCD sum rules, potential models (for quarkonium) etc. But the perturbative approach, based on asymptotic freedom and only applicable to hard processes, still remains the main quantitative connection to experiment. All of this is very important for the LHC preparation: understanding QCD processes is an essential prerequisite for all possible discoveries. Great experimental work on testing QCD has been accomplished over the years. In this respect it is very appropriate to pay here a tribute to HERA that has done a wonderful job in this domain. Great results are still coming out from HERA experiments \cite{AGlazov, MRuspa}, like the measurements of the longitudinal structure function, of the diffractive structure functions, of the contribution of heavy quarks and so on. New interesting results have been found in heavy flavour spectroscopy \cite{JBrodzicka}. Measurements of QCD processes at the Tevatron have also been of the utmost importance \cite{MWobisch}.

Due to confinement no free coloured particles are observed but only colour singlet hadrons. In high energy collisions the produced quarks and gluons materialize as narrow jets of hadrons. Our understanding of the confinement mechanism has much improved thanks to lattice simulations of QCD at finite temperatures and densities \cite{AUkawa}.  The potential between two colour charges, obtained from the lattice computations, clearly shows a linear slope at large distances (linearly rising potential). The slope decreases with increasing temperature until it vanishes at a critical temperature $T_C$. Above $T_C$ the slope remains zero. The phase transitions of colour deconfinement and of chiral restoration appear to happen together on the lattice. Near the critical temperature for both deconfinement and chiral restoration a rapid transition is observed in lattice simulations. In particular the energy density $\epsilon(T)$ is seen to sharply increase. The critical parameters and the nature of the phase transition depend on the number of quark flavours $N_f$ and on their masses. For example, for  $N_f$ = 2 or 2+1 (i.e. 2 light u and d quarks and 1 heavier s quark), $T_C \sim 175~MeV$  and $\epsilon(T_C) \sim 0.5-1.0 ~GeV/fm^3$. For realistic values of the masses $m_s$ and $m_{u,d}$ the phase transition appears to be a second order one, while it becomes first order for very small or very large $m_{u,d,s}$. At high densities the colour superconducting phase is probably also present with diquarks acting as Cooper pairs. The hadronic phase and the deconfined phase are separated by a crossover line at small densities and by a critical line at high densities. Determining the exact location of the critical point in T and $\mu_B$ is an important challenge for theory and is also important for the interpretation of heavy ion collision experiments.  

A large investment is being done in experiments of heavy ion collisions with the aim of finding some evidence of the quark gluon plasma phase. Many exciting results have been found at the CERN SPS in the past  years and more recently at RHIC \cite{TPeitzmann}. At the CERN SPS some experimental hints of rapid variation of measured quantities with the energy density were found in the form, for example, of $J/ \Psi$ production suppression or of strangeness enhancement when going from p-A to Pb-Pb collisions. Indeed a posteriori the CERN SPS appears well positioned in energy to probe the transition region, in that a marked variation of different observables was observed. One impressive effect detected at RHIC, interpreted as due to the formation of a hot and dense bubble of matter, is the observation of a strong suppression of back-to-back correlations in jets from central collisions in Au-Au, showing that the jet that crosses the bulk of the dense region is absorbed. The produced hot matter shows a high degree of collectivity \cite{TRenk}, as shown by the observation of elliptic flow (produced hadrons show an elliptic distribution while it would be spherical for a gas) and resembles a perfect liquid with small or no viscosity. There is also evidence for a 2-component hadronisation mechanism: 
coalescence \cite{ref:coa} and fragmentation. Early produced partons with high density show an exponential falling in $p_T$: they produce hadrons by joining together. At large $p_T$ fragmentation with power behaviour survives. Elliptic flow, inclusive spectra, partonic energy loss in medium, strangeness enhancement, J/$\Psi$  suppression etc. are all suggestive (but only suggestive!) of early production of a coloured partonic medium with high energy density
and temperature, close to the theoretically expected values, then expanding as a near ideal fluid. The experimental programme on heavy ion collisions will continue at the LHC where ALICE, the dedicated heavy ion collision experiment, is ready to take data \cite{PBraun-Munzinger}.

As we have seen, a main approach to non perturbative problems in QCD is by simulations of the theory on the lattice \cite{AUkawa}, a technique started by K. Wilson in 1974 which has shown continuous progress over the last decades by going to smaller lattice spacing and larger lattices. A recent big step, made possible by the availability of more powerful dedicated computers, is the evolution from quenched (i.e. with no dynamical fermions) to unquenched calculations. Calculations with dynamical fermions (which take into account the effects of virtual quark loops) imply the evaluation of the quark determinant which is a difficult task. How difficult depends on the particular calculation method. There are several approaches (Wilson, twisted mass,  Kogut-Susskind staggered, Ginsparg-Wilson fermions), each with its own advantages and disadvantages (including the time it takes to run the simulation on a computer). Another important progress is in the capability of doing the simulations with lighter quark masses (closer to the physical mass).  As lattice simulations are always limited to masses of light quarks larger than a given value, going to lighter quark masses makes the use of chiral extrapolations less important (to extrapolate the results down to the physical pion mass one can take advantage of the chiral effective theory in order to control the chiral logs: $\log(m_q/4\pi f_\pi)$).  With the progress from unquenching and lighter quark masses an evident improvement in the agreement of predictions with the data is obtained. For example, modern simulations reproduce the hadron spectrum quite well.  For lattice QCD one is now in an epoch of pre-dictivity as opposed to the  post-dictivity of the past. And in fact the range of precise lattice results currently includes many domains:  the QCD coupling constant (the value $\alpha_s(m_Z)=0.1184(4)$ has been quoted \cite{AUkawa}: the central value is in agreement with other determinations but I would not trust the stated error as a fair representation of the total uncertainty), the quark masses, the form factors for K and D decay, the B parameter for kaons, the decay constants $f_K$, $f_D$, $f_{Ds}$, the $B_c$ mass and many more.

We  now discuss  perturbative QCD. In the QCD Lagrangian quark masses are the only parameters with dimensions. Naively (or classically) one would expect massless QCD to be scale invariant so that dimensionless observables would not depend on the absolute energy scale but only on ratios of energy variables. While massless QCD in the quantum version, after regularisation and renormalisation, is finally not scale invariant, the theory is asymptotically free and all the departures from scaling are asymptotically small and computable in terms of the running coupling $\alpha_s(Q^2)$ that decreases logarithmically at large $Q^2$. Mass corrections, present in the realistic case together with other non perturbative effects, are suppressed by powers of $1/Q^2$. 

The measurements of $\alpha_s(Q^2)$  are among the main quantitative tests of the theory. The most  precise and  reliable determinations are from $e^+e^-$ colliders (mainly at LEP: inclusive Z decays, inclusive
hadronic $\tau$ decay, event shapes and jet rates) and from scaling violations in Deep Inelastic Scattering (DIS).  There is a remarkable agreement among these different determinations. An all-inclusive average $\alpha_s(m_Z^2)=0.1184(7)$ is obtained in \cite{beth}, a value which corresponds to $\Lambda_{QCD}\sim 213(9)~MeV$ ($\bar{MS}$, 5 flavours).

Since $\alpha_s$ is not too small, $\alpha_s(m_Z^2) \sim 0.12$, the need of high
order perturbative calculations, of resummation of logs at all 
orders etc. is particularly acute.  Ingenious new computational techniques and software have been developed and many calculations have been realized that only a decade ago appeared as impossible \cite{NGlover}.  An increasing number of processes of interest for the physics at the LHC have been computed at NLO. Recent examples are the NLO calculations for $q \bar q \rightarrow t \bar t b \bar b$ \cite{pozzo} and for $W \rightarrow$ 3 jets \cite{w3j}. Methods for the automated calculation of NLO processes have been very much advanced, based on generalised unitarity \cite{uni} and algebraic reduction to basic integrals at the integrand level \cite{opp}.  Powerful tools have been developed for automatic NLO calculations like HELAC, CutTools, BlackHat, Rocket \cite{pit}.

Important work on jet recombination algorithms has been published by G. Salam and collaborators (for a review, see \cite{sal1}). In fact it is essential that a correct jet finding is implemented by LHC experiments for an optimal matching of theory and experiment. A critical reappraisal of the existing cone and recombination methods has led to new improved versions of jet defining algorithms, like SISCone \cite{sal2} and anti-$k_T$ \cite{sal3}, with good infra red properties and leading to a simpler jet structure.

For benchmark measurements where experimental errors are small and corrections are large NNLO calculations are needed. A number of these extremely sophisticated calculations have been completed. 
In 2004 the complete calculation of the NNLO splitting functions has been published \cite{ref:moc} $\alpha_s P \sim \alpha_s P_1+ \alpha_s^2 P_2 + \alpha_s^3 P_3+\dots$, a really monumental, fully analytic, computation. More recently the main part of the inclusive hadronic $Z$ and $\tau$ decays at $o(\alpha_s^4)$ (NNNLO!) has been computed \cite{ref:bck}. The calculation (which involves  some 20.000 diagrams) is complete for $\tau$ decay, while for $Z$ decay only the non singlet terms, proportional to $\Sigma_f Q_f^2$, are included
( but singlet terms ~$(\Sigma_f Q_f)^2)$ are small at the previous order $o(\alpha_s^3)$). 
The calculation of the hadronic event shapes in $e^+e^-$ annihilation at NNLO has also been completed \cite{ref:ggh}, which involves consideration of 3, 4 and 5 jets with one loop corrections for 4 jets and two loop corrections for 3 jets. 
These calculations were applied in ref.\cite{ref:dis} to the measurement of $\alpha_s$ from data on event shapes obtained by ALEPH with the result $\alpha_s(m_Z^2)=0.1224\pm 0.0039$. 

Another very important example is Higgs production via $g ~+~ g \rightarrow H$ \cite{ref:boz}. The amplitude is dominated by the top quark loop (if heavier coloured particles exist they also would contribute). The NLO corrections turn out to be particularly large. Higher order corrections can be computed either in the effective lagrangian approach, where the heavy top is integrated away and the loop is shrunk down to a point (the coefficient of the effective vertex is known to $\alpha_s^4$ accuracy), or in the full theory. At the NLO the two approaches agree very well for the rate as a function of $m_H$. The NNLO corrections have been computed in the effective vertex approximation. Beyond fixed order, resummation of large logs were carried out. Also the NLO EW contributions are known by now. Rapidity (at  NNLO) and $p_T$ distributions (at NLO) have also been evaluated. At smaller $p_T$ the large logarithms $[log(p_T/m_H)]^n$ have been resummed in analogy with what was done long ago for W and Z production. 

The importance of DIS for QCD goes well beyond the measurement of $\alpha_s$. In the past it played a crucial role in establishing the reality of quarks and gluons as partons and in promoting  QCD as the theory of strong interactions. Nowadays it still generates challenges to QCD as, for example, in the domain of structure functions at small x or of polarized structure functions or of generalized parton densities and so on. 

The problem of constructing a convergent procedure to include the BFKL corrections at small x in the singlet splitting functions, in agreement with the small-x behaviour observed at HERA, has been a long standing puzzle which has now been essentially solved. The naive BFKL rise of splitting functions is tamed by resummation of collinear singularities and by running coupling effects. The resummed expansion is well behaved and the result is close to the perturbative NLO splitting function in the region of HERA data at small x \cite{ref:abf},\cite{ref:ccss}. 

In polarized DIS one main question is how the proton helicity is distributed among quarks, gluons and orbital angular momentum: $1/2\Delta \Sigma + \Delta g + L_z= 1/2$ \cite{DHasch}.
The quark moment $\Delta \Sigma$ was found to be small: typically, at $Q^2\sim 1~GeV^2$, $\Delta \Sigma_{exp} \sim 0.3$ (the "spin crisis") \cite{ref:spi1}. Either $\Delta g + L_z$ is large or there are contributions to $\Delta \Sigma$ at very small x outside of the measured region. $\Delta g$ evolves like $\Delta g \sim log Q^2$, so that eventually should become large (while $\Delta \Sigma$ and $\Delta g + L_z$ are $Q^2$ independent in LO). For conserved quantities we would expect that they are the same for constituent and for parton quarks. But actually the conservation of $\Delta \Sigma$ is broken by the axial anomaly and, in fact, in perturbation theory beyond LO the conserved density is actually $\Delta \Sigma'=\Delta \Sigma+n_f/2\pi \alpha_s~\Delta g$  \cite{ref:spi1}. Note that also $\alpha_s \Delta g$ is conserved in LO, as $\Delta g \sim \log{Q^2}$. This behaviour is not controversial but it will take long before the log growth of $\Delta g$ will be confirmed by experiment! But by establishing this behaviour  one would show that the extraction of $\Delta g$ from the data is correct and that the QCD evolution works as expected.  If $\Delta g$ was large enough it could account for the difference between partons ($\Delta \Sigma$) and constituents ( $\Delta \Sigma'$). From the spin sum rule it is clear that the log increase should cancel between $\Delta g$ and  $L_z$. This cancelation is automatic as a consequence of helicity conservation in the basic QCD vertices.  Existing direct measurements by Hermes, Compass, and at RHIC are still very crude and show no hint of a large $\Delta g$ \cite{ref:spi2} at accessible values of $x$ and $Q^2$.  Present data are consistent with $\Delta g$ large enough to sizeably contribute to the spin sum rule but there is no indication that $\alpha_s \Delta g$ can explain the difference between constituents and parton quarks. 

Another important role of DIS is to provide information on parton density functions (PDF) \cite{heralhc} which are instrumental for computing cross-sections of hard processes at hadron colliders via the factorisation formula. The predictions for cross sections and distributions at $pp$ or $p\bar p$ colliders for large $p_T$ jets or photons, for heavy quark production, for Drell-Yan, W and Z production are all in very good agreement with experiment. There was an apparent problem for b quark production at the Tevatron, but the problem appears now to be solved by a combination of refinements (log resummation, B hadrons instead of b quarks, better fragmentation functions....)\cite{ref:cac}. The QCD predictions are so solid that W and Z production are actually considered as possible luminosity monitors for the LHC. 

The activity on event simulation also received a big boost from the LHC preparation (see, for example, \cite{am} and the review \cite{ref:LHC}). General algorithms for performing NLO calculations numerically (requiring techniques for the cancellation of singularities between real and virtual diagrams) have been developed (see, for example, \cite{ref:num}). The matching of matrix element calculation of rates together with the modeling of parton showers has been realised in packages, as for example in the MC@NLO \cite{ref:frix} or POWHEG \cite{ref:fnr} based on HERWIG. The matrix element calculation, improved by resummation of large logs, provides the hard skeleton (with large $p_T$ branchings) while the parton shower is constructed by a sequence of factorized collinear emissions fixed by the QCD splitting functions. In addition, at low scales a model of hadronisation completes the simulation. The importance of all the components, matrix element, parton shower and hadronisation can be appreciated in simulations of hard events compared with the Tevatron data. 

In conclusion, I think that the domain of QCD appears as one of great maturity but also of robust vitality (as apparent by the large amount of work produced for the LHC preparation) and all the QCD predictions that one was able to formulate and to test are in very good agreement with experiment.

\section{The Higgs Problem}

The Higgs problem is really central in particle physics today \cite{wells}. On the one hand, the experimental verification of the Standard Model (SM) cannot be considered complete until the structure of the  Higgs sector is not established by experiment. On the other hand, the Higgs is directly related to most of the major open problems of particle physics, like the flavour problem and the hierarchy problem, the latter strongly suggesting the need for new physics near the weak scale, which could also clarify the dark matter identity. It is clear that the fact that some sort of Higgs mechanism is at work has already been established. The longitudinal degree of freedom for the W or the Z is borrowed from the Higgs sector and is an evidence for it. In fact the couplings of quarks and leptons to
the weak gauge bosons W$^{\pm}$ and Z are indeed precisely those
prescribed by the gauge symmetry.  To a lesser
accuracy the triple gauge vertices $\gamma$WW and ZWW have also
been found in agreement with the specific predictions of the
$SU(2)\bigotimes U(1)$ gauge theory. This means that it has been
verified that the gauge symmetry is unbroken in the vertices of the
theory: all currents and charges are indeed symmetric. Yet there is obvious
evidence that the symmetry is instead badly broken in the
masses. The W or the Z with longitudinal polarization that are observed are not present in an unbroken gauge theory (massless spin-1 particles, like the photon, are transversely polarized). Not only the W and the Z have large masses, but the large splitting of, for example,  the t-b doublet shows that even the global weak SU(2) is not at all respected by the fermion spectrum. Symmetric couplings and totally non symmetric spectrum is a clear signal of spontaneous
symmetry breaking and its implementation in a gauge theory is via the Higgs mechanism. The big remaining questions are about the nature and the properties of the Higgs particle(s). 

The LHC has been designed to solve the Higgs problem. A strong argument indicating that the solution of the Higgs problem cannot be too far away is the fact that, in the absence of a Higgs particle or of an alternative mechanism, violations of unitarity appear in scattering amplitudes involving longitudinal gauge bosons (those most directly related to the Higgs sector) at energies in the few TeV range \cite{ref:unit}. A crucial question for the LHC is to identify the mechanism that avoids the unitarity violation: is it one or more Higgs bosons or some new vector boson (like additional gauge bosons WÕ, ZÕ or Kaluza-Klein recurrences or resonances from a strong sector) \cite{CGrojean, SPokorski}?

It is well known that in the SM with only one Higgs doublet a lower limit on
$m_H$ can be derived from the requirement of vacuum stability (i.e. that the quartic Higgs coupling $\lambda$ does not turn negative in its running up to a large scale $\Lambda$) or, in milder form, of a moderate instability, compatible with the lifetime of the Universe  \cite{ref:isid}. The Higgs mass enters because it fixes the initial value of the quartic Higgs coupling $\lambda$. For the experimental value of $m_t$ the lower limit is below the direct experimental bound for $\Lambda \sim $ a few TeV and is $M_H> 130$ GeV for $\Lambda \sim M_{Pl}$. Similarly an upper bound on $m_H$ (with mild dependence
on $m_t$) is obtained, as described in \cite{ref:hri}, from the requirement that for $\lambda$ no Landau pole appears up to the scale $\Lambda$, or in simpler terms, that the perturbative description of the theory remains valid up to  $\Lambda$. The upper limit on the Higgs mass in the SM is clearly important for assessing the chances of success of the LHC as an accelerator designed to solve the Higgs problem. Even if $\Lambda$ is as small as ~a few TeV the limit is $m_H < 600-800~$GeV and becomes $m_H < 180~$GeV for $\Lambda \sim M_{Pl}$. 

In conclusion it looks very likely that the LHC can clarify the problem of the electroweak symmetry breaking mechanism. It has been designed for it!

\section{Precision Tests of the Standard Electroweak Theory}

The most precise tests of the electroweak theory apply to the QED sector. The anomalous magnetic moments of the electron and of the muon are among the most precise measurements in the whole of physics \cite{TMori}, \cite{JJaeckel}. Recently there have been new precise measurements of $a$ for the electron \cite{ref:ae1} and the muon \cite{ref:amu} ($a = (g-2)/2$). The QED part has been computed analytically for $i=1,2,3$, while for $i=4$ there is a numerical calculation (see, for example, \cite{ref:kino}). Some terms for $i=5$ have also been estimated for the muon case. The weak contribution is from $W$ or $Z$ exchange. The hadronic contribution is from vacuum polarization insertions and from light by light scattering diagrams.  For the electron case the weak contribution is essentially negligible and the hadronic term does not introduce an important uncertainty.  As a result the $a_e$ measurement can be used to obtain the most precise determination of the fine structure constant \cite{ref:ae2}. In the muon case the experimental precision is less by about 3 orders of magnitude, but the sensitivity to new physics effects is typically increased by a factor $(m_\mu/m_e)^2 \sim 4^.10^4$. The dominant theoretical ambiguities arise from the hadronic terms in vacuum polarization and in light by light scattering. If the vacuum polarization terms are evaluated from the $e^+e^-$ data a discrepancy of $\sim 3 \sigma$ is obtained (the $\tau$ data would indicate better agreement, but the connection to $a_\mu$ is less direct and recent new data have added solidity to the $e^+e^-$ route)\cite{ref:amu2}. Finally, we note that, given the great accuracy of the $a_\mu$ measurement and the estimated size of the new physics contributions, for example from SUSY, it is not unreasonable that a first signal of new physics would appear in this quantity.

The results of the electroweak precision tests also including the measurements of $m_t$, $m_W$ and the searches for new physics at the Tevatron \cite{FCanelli} form a very stringent set of precise constraints \cite{ref:ewg} to compare with the Standard Model (SM) or with
any of its conceivable extensions \cite{SDittmaier}. When confronted with these results, on the whole the SM performs rather
well, so that it is fair to say that no clear indication for new physics emerges from the data \cite{ref:AG}.  But the
Higgs sector of the SM is still very much untested. What has been
tested is the relation $M_W^2=M_Z^2\cos^2{\theta_W}$, modified by small, computable
radiative corrections. This relation means that the effective Higgs
(be it fundamental or composite) is indeed a weak isospin doublet.
The Higgs particle has not been found but in the SM its mass can well
be larger than the present direct lower limit $m_H > 114.4$~GeV
obtained from direct searches at LEP-2.  The radiative corrections
computed in the SM when compared to the data on precision electroweak
tests lead to a clear indication for a light Higgs, not too far from
the present lower bound. The exact upper limit for $m_H$ in the SM depends on the value of the top quark mass $m_t$ (the one-loop radiative corrections are quadratic in $m_t$ and logarithmic in $m_H$). The measured value of $m_t$ went down recently (as well as the associated error) according to the results of Run II at the Tevatron. The CDF and D0 combined value is at present $m_t~= 173.1~\pm~1.3~GeV$. As a consequence the present limit on $m_H$ is quite stringent: $m_H < 186~GeV$ (at $95\%$ c.l., after including the information from the 114.4 GeV direct bound)  \cite{ref:ewg}.  

In the Higgs search the Tevatron is now reaching the SM sensitivity. At this Symposium the quoted result for the SM Higgs is that the interval $160 < m_H < 170$ GeV is excluded at $95\%$ c.l. \cite{GBernardi}. But the most recent limit, reported near the end of 2009, is somewhat weaker: $163 < m_H < 166$ GeV \cite{SJindariani}. The goal at Fermilab is to collect $12~fb^{-1}$ of luminosity by 2011 and possibly exclude $115 < m_H < 185$ GeV.

\section{The Physics of Flavour}

Another domain where the SM is really in good agreement with the data is flavour physics (actually too good in comparison with the general expectation before the experiments). In the last decade great progress in different areas of flavour physics has been achieved. In the quark sector, the amazing results of a generation of frontier experiments, performed at B factories and at accelerators, have become available. QCD has been playing a crucial role in the interpretation of experiments by a combination of effective theory methods (heavy quark effective theory, NRQCD, SCET), lattice simulations and perturbative calculations. 
The hope of the B-decay experiments was to detect departures from the CKM picture of mixing and of CP violation as  signals of new physics. At present the available results on B mixing and CP violation on the whole agree very well with the SM predictions based on the CKM matrix \cite{SPrell}, \cite{TKomatsubara}, \cite{rept}. A few interesting ÒtensionsÓ at the 2-3 $\sigma$ level should be monitored closely in the future \cite{SPrell}, \cite{TIijima}: $\sin{2\beta}$ from $B_d \rightarrow J/\Psi K^0$ versus $\epsilon_K$ and $V_{ub}$ (which, however, in my opinion, is probably due to an underestimate of theoretical errors, particularly on the determination of $V_{ub}$), $\beta_s$ measured by CDF, $D_0$
in $B_s \rightarrow J/\Psi \phi$ and $B\rightarrow \tau \nu$. But certainly the amazing performance of the SM in flavour changing  and/or CP violating transitions in K and B decays poses very strong constraints on all proposed models of new physics \cite{GHiller}, \cite{isid}. For example, if one adds to the SM effective non renormalizable operators suppressed by powers of a scale $\Lambda$ one generally finds that experiments indicate very large values of $\Lambda$, much above the few TeV range indicated by the hierarchy problem. Only if one assumes that the deviations from new physics occur at loop level and inherit the same SM protections against flavour changing neutral currents (like the GIM mechanism and small $V_{CKM}$ factors) as, for example, in Minimal Flavour Violation models \cite{isid},  that one obtains bounds on  $\Lambda$ in the few TeV range.

In the leptonic sector the study of neutrino oscillations has led to the discovery that at least two neutrinos are not massless and to the determination of the mixing matrix \cite{JValle}, \cite{revnu}. Neutrinos are not all massless but their masses are very small (at most a fraction of $eV$). The neutrino spectrum could be either of the normal hierarchy type (with the solar doublet below), or of the inverse hierarchy type (with the solar doublet above). Probably masses are small because $\nu$Õs are Majorana fermions, and, by the see-saw mechanism, their masses are inversely proportional to the large scale $M$ where lepton number ($L$) non conservation occurs (as expected in GUT's). Indeed the value of $M\sim m_{\nu R}$ from experiment is compatible with being close to $M_{GUT} \sim 10^{14}-10^{15}GeV$, so that neutrino masses fit well in the GUT picture and actually support it. The interpretation of neutrinos as Majorana particles enhances the importance of experiments aimed at the detection of neutrinoless double beta decay and a huge effort in this direction is underway  \cite{CWeinheimer}.  It was realized that decays of heavy $\nu_R$ with CP and L non conservation can produce a B-L asymmetry (which is unchanged by instanton effects at the electroweak scale). The range of neutrino masses indicated by neutrino phenomenology turns out to be perfectly compatible with the idea of baryogenesis via leptogenesis \cite{ref:buch}. This elegant model for baryogenesis has by now replaced the idea of baryogenesis near the weak scale, which has been strongly disfavoured by LEP. It is remarkable that we now know the neutrino mixing matrix with good accuracy \cite{datanu}. Two mixing angles are large and one is small. The atmospheric angle $\theta_{23}$ is large, actually compatible with maximal but not necessarily so. The solar angle $\theta_{12}$ (the best measured) is large, $\sin^2{\theta_{12}}\sim 0.3$, but certainly not maximal. The third angle $\theta_{13}$, strongly limited mainly by the CHOOZ experiment, has at present a $3\sigma$ upper limit given by about $\sin^2{\theta_{13}}\leq 0.04$. It is a fact that, to a precision comparable with the measurement accuracy, the Tri-Bimaximal (TB) mixing pattern ($\sin^2{\theta_{12}}\sim 1/3$, $\sin^2{\theta_{23}}\sim 1/2$ and $\sin^2{\theta_{13}} = 0$) \cite{Harr} is well approximated by the data. If this experimental result is not a mere accident but a real indication that a dynamical mechanism is at work to guarantee the validity of TB mixing in the leading approximation, corrected by small non leading terms, then non abelian discrete flavor groups emerge as the main road to an understanding of this mixing pattern \cite{rmp}. Indeed the entries of the TB mixing matrix are clearly suggestive of "rotations" by simple, very specific angles. In fact the group $A_4$, the simplest group used to explain TB mixing, is defined as the group of rotations that leave a regular rigid tetrahedron invariant. The non conservation of the three separate lepton numbers and the large leptonic mixing angles make it possible that processes like $\mu \rightarrow e \gamma$ or $\tau \rightarrow \mu \gamma$ might be observable, not in the SM but in extensions of it like the MSSM. Thus, for example, the outcome of the now running experiment MEG at PSI \cite{TMori} aiming at improving the limit on $\mu \rightarrow e \gamma$ by 1 or 2 orders of magnitude, is of great interest. 

\section{Outlook on Avenues beyond the Standard Model}

No signal of new physics has been
found neither in electroweak precision tests nor in flavour physics \cite{OGLopez}. Given the success of the SM why are we not satisfied with that theory? Why not just find the Higgs particle,
for completeness, and declare that particle physics is closed? The reason is that there are
both conceptual problems and phenomenological indications for physics beyond the SM. On the conceptual side the most
obvious problems are the proliferation of parameters, the puzzles of family replication and of flavour hierarchies, the fact that quantum gravity is not included in the SM and the related hierarchy problem. Among the main
phenomenological hints for new physics we can list dark matter, the quest for Grand Unification and coupling constant merging, neutrino masses (explained in terms of L non conservation), 
baryogenesis and the cosmological vacuum energy (a gigantic naturalness problem).

We know by now \cite{BAtwood}, \cite{NSmith}, \cite{KOlive} that  the  Universe is flat and most of it is not made up of known forms of matter: while $\Omega_{tot} \sim 1$ and $\Omega_{matter} \sim 0.3$, the normal baryonic matter is only $\Omega_{baryonic} \sim 0.044$, where $\Omega$ is the ratio of the density to the critical density. Most of the energy in the Universe is dark matter (DM) and Dark Energy (DE) with $\Omega_{\Lambda} \sim 0.7$. We also know that most of DM must be cold (non relativistic at freeze-out) and that significant fractions of hot DM are excluded. Neutrinos are hot DM (because they are ultrarelativistic at freeze-out) and indeed are not much cosmo-relevant: $\Omega_{\nu} \lappeq 0.015$. The identification of DM is a task of enormous importance for both particle physics and cosmology. The LHC has good chances to solve this problem in that it is sensitive to a large variety of WIMP's (Weekly Interacting Massive Particles). WIMP's with masses in the 10 GeV-1 TeV range with typical EW cross-sections turn out to contribute terms of $o(1)$ to $\Omega$. This is a formidable hint in favour of WIMP's as DM candidates. By comparison, axions are also DM candidates but their mass and couplings must be tuned for this purpose. If really some sort of WIMP's are a main component of DM they could be discovered at the LHC and this will be a great service of particle physics to cosmology. Active searches in non-accelerator experiments are under way  \cite{NSmith}. Some hints of possible signals have been reported: e.g. annual modulations (DAMA/LIBRA at Gran Sasso \cite{ber}), $e^+$ and/or $e^+e^-$ excess in cosmic ray detectors, e.g. in PAMELA \cite{pam} and ATIC \cite{atic} (but the ATIC excess has not been confirmed by Fermi-LAT \cite{fer}). If those effects are really signals for DM they would indicate particularly exotic forms of DM \cite{exdm}. But for the PAMELA effect an astrophysical explanation in terms of relatively close pulsars appears as a plausible alternative \cite{BAtwood}.

The computed evolution with energy
of the effective gauge couplings clearly points towards the unification of the electro-weak and strong forces (Grand Unified Theories: GUT's) at scales of energy $M_{GUT}\sim  10^{15}-10^{16}~ GeV$ which are close to the scale of quantum gravity, $M_{Pl}\sim 10^{19}~ GeV$.  One is led to imagine  a unified theory of all interactions also including gravity (at present superstrings provide the best attempt at such a theory \cite{SKachru}). Thus GUT's and the realm of quantum gravity set a very distant energy horizon that modern particle theory cannot ignore. Can the SM without new physics be valid up to such large energies? One can imagine that some of the obvious problems of the SM could be postponed to the more fundamental theory at the Planck mass. For example, the explanation of the three generations of fermions and the understanding of the pattern of fermion masses and mixing angles can be postponed. But other problems must find their solution in the low energy theory. In particular, the structure of the SM could not naturally explain the relative smallness of the weak scale of mass, set by the Higgs mechanism at $\mu\sim 1/\sqrt{G_F}\sim  250~ GeV$  with $G_F$ being the Fermi coupling constant. This so-called hierarchy problem \cite{CGrojean}, \cite{SPokorski} is due to the instability of the SM with respect to quantum corrections. This is related to the presence of fundamental scalar fields in the theory with quadratic mass divergences and no protective extra symmetry at $\mu=0$. For fermion masses, first, the divergences are logarithmic and, second, they are forbidden by the $SU(2)\bigotimes U(1)$ gauge symmetry plus the fact that at $m=0$ an additional symmetry, i.e. chiral  symmetry, is restored. Here, when talking of divergences, we are not worried of actual infinities. The theory is renormalizable and finite once the dependence on the cut-off $\Lambda$ is absorbed in a redefinition of masses and couplings. Rather the hierarchy problem is one of naturalness. We can look at the
cut-off as a parameterization of our ignorance on the new physics that will modify the theory at large energy
scales. Then it is relevant to look at the dependence of physical quantities on the cut-off and to demand that no unexplained enormously accurate cancellations arise. 
In fact, the hierarchy problem can be put in quantitative terms: loop corrections to the higgs mass squared are quadratic in the cut-off $\Lambda$. The most pressing problem is from the top loop.
 With $m_h^2=m^2_{bare}+\delta m_h^2$ the top loop gives 
 \begin{eqnarray}
\delta m_{h|top}^2\sim -\frac{3G_F}{2\sqrt{2} \pi^2} m_t^2 \Lambda^2\sim -(0.2\Lambda)^2 \label{top}
\end{eqnarray}
If we demand that the correction does not exceed the light Higgs mass indicated by the precision tests, $\Lambda$ must be
close, $\Lambda\sim o(1~TeV)$. Similar constraints also arise from the quadratic $\Lambda$ dependence of loops with gauge bosons and
scalars, which, however, lead to less pressing bounds. So the hierarchy problem demands new physics to be very close (in
particular the mechanism that quenches the top loop). Actually, this new physics must be rather special, because it must be
very close, yet its effects are not clearly visible in precision electroweak tests - the "LEP Paradox" \cite{ref:BS} - now also accompanied by a similar "flavour paradox" \cite{isid}. Examples  \cite{CGrojean}, \cite{SPokorski} of proposed classes of solutions for the hierarchy problem are :

¥ $\bf{Supersymmetry.}$ In the limit of exact boson-fermion symmetry \cite{ref:Martin} the quadratic divergences of bosons cancel so that
only log divergences remain. However, exact SUSY is clearly unrealistic. For approximate SUSY (with soft breaking terms),
which is the basis for all practical models, $\Lambda$ is replaced by the splitting of SUSY multiplets, $\Lambda\sim
m_{SUSY}-m_{ord}$. In particular, the top loop is quenched by partial cancellation with s-top exchange, so the s-top cannot be too heavy. An important phenomenological indication is that coupling unification is not exact in the SM while it is quantitatively precise in SUSY GUT's where also proton decay bounds are not in contradiction with the predictions. An interesting exercise is to repeat the fit of precision tests in the Minimal Supersymmetric Standard Model with GUT constraints added, also including the additional data on the muon $(g-2)$, the dark matter relic density and on the $b\rightarrow s \gamma$ rate. The result is that the central value of the lightest Higgs mass $m_h$ goes up (in better harmony with the bound from direct searches) for moderately large $tan\beta$ and relatively light SUSY spectrum \cite{ref:sus}.

¥ $\bf{Technicolor.}$ The Higgs system is a condensate of new fermions. There are no fundamental scalar Higgs sector, hence no
quadratic devergences associated to the $\mu^2$ mass in the scalar potential. This mechanism needs a very strong binding force,
$\Lambda_{TC}\sim 10^3~\Lambda_{QCD}$. It is  difficult to arrange that such nearby strong force is not showing up in
precision tests. Hence this class of models has been disfavoured by LEP, although some special class of models have been devised a posteriori, like walking TC, top-color assisted TC etc \cite{ref:L-C}. 

¥ $\bf{Extra~dimensions.}$ One possibility is that $M_{Pl}$ appears very large, or equivalently that gravity appears very weak,
because we are fooled by hidden extra dimensions so that the real gravity scale is reduced down to a lower scale, even possibly down to
$o(1~TeV)$ ("large" extra dimensions). This possibility is very exciting in itself and it is really remarkable that it is not directly incompatible with experiment but a realistic model has not emerged \cite{ref:Jo}. In fact, the most promising set of extra dimensional models are those with "warped" metric, which offer attractive solutions to the hierarchy problem \cite{ref:RS,ref:sura}. An important direction of development is the study of symmetry breaking by orbifolding and/or boundary conditions. These are models where a larger gauge symmetry (with or without SUSY) holds in the bulk. The symmetry is reduced on the 4 dimensional brane, where the physics that we observe is located, as an effect of symmetry breaking induced geometrically by suitable boundary conditions (see, for example, the class of models in \cite{ref:co}). Also "Higgsless  models" have been tried where it is the SM electroweak gauge symmetry which is broken at the boundaries \cite{ref:Hless} (then no Higgs should be found at the LHC but other signals, like additional vector bosons, should appear). Extra dimensions offer a rich and exciting general framework.

¥ $\bf{"Little~Higgs"~models.}$ In these models extra symmetries allow $m_h\not= 0$ only at two-loop level, so that $\Lambda$
can be as large as
$o(10~TeV)$ with the Higgs within present bounds (the top loop is quenched by exchange of heavy vectorlike new  quarks with charge 2/3) \cite{ref:schm}. Certainly these models involve a remarkable level of group theoretic virtuosity. However, in
the simplest versions one is faced with problems with precision tests of the SM. These bad features can be fixed by some suitable complication of the model (see for example, \cite{ref:Ch}). But, in my opinion, the real limit of
this approach is that it only offers a postponement of the main problem by a few TeV, paid by a complete loss of
predictivity at higher energies. In particular all connections to GUT's are lost.

¥ $\bf{Effective~theories~for~compositeness.}$  In this approach \cite{ref:comp} a low energy theory, left over by truncation of some UV completion, is described in terms of an elementary sector (the SM particles minus the Higgs) a composite sector (including the Higgs, massive vector bosons $\rho_\mu$ and new fermions) and a mixing sector. The Higgs is a pseudo Goldstone bosons of a larger broken gauge group, with $\rho_\mu$ the corresponding massive vector bosons. Mass eigenstates are mixtures of elementary and composite states, with light particles mostly elementary and heavy particles mostly composite. But the Higgs is totally composite (perhaps also the right-handed top quark). New physics in the composite sector is well hidden because light particles have small mixing angles. The Higgs is light because only acquires
mass through interactions with the light particles from their composite components. This general description can apply to models with a strongly interacting sector as arising from little Higgs or extra dimension scenarios.

¥ $\bf{The~anthropic~solution.}$ The apparent value of the cosmological constant $\Lambda$ poses a tremendous, unsolved naturalness problem \cite{ref:tu}. Yet the value of $\Lambda$ is close to the Weinberg upper bound for galaxy formation \cite{ref:We}. Possibly our Universe is just one of infinitely many (Multiverse) continuously created from the vacuum by quantum fluctuations. Different types of physics are realized in different Universes according to the multitude of string theory solutions (~$10^{500}$). Perhaps we live in a very unlikely Universe but the only one that allows our existence \cite{ref:anto}. I find applying the anthropic principle to the SM hierarchy problem excessive. After all we can find plenty of models that easily reduce the fine tuning from $10^{14}$ to $10^2$: why make our Universe so terribly unlikely? By comparison the case of the cosmological constant is a lot different: the context is not as fully specified as the for the SM (quantum gravity, string cosmology, branes in extra dimensions, wormholes through different Universes....).

From model building the following lessons can be derived. In all the new physics models we have mentioned
there is a light Higgs ($\lappeq$ 200 GeV), except in Higgsless models (if any) but new
light new vector bosons exist in this case. In all models there is at least a percent fine tuning, so that
fine tuning appears to be imposed on us by the data.

\section{Conclusion}

Supersymmetry remains the standard way beyond the SM. What is unique to SUSY, beyond leading to a set of consistent and
completely formulated models, as, for example, the MSSM, is that this theory can potentially work up to the GUT energy scale.
In this respect it is the most ambitious model because it describes a computable framework that could be valid all the way
up to the vicinity of the Planck mass. The SUSY models are perfectly compatible with GUT's and are actually quantitatively
supported by coupling unification and compatible with proton decay bounds and also by what we have recently learned on neutrino masses. All other main ideas for going
beyond the SM do not share this synthesis with GUT's. The SUSY way is testable, for example at the LHC, and the issue
of its validity will be decided by experiment. It is true that we could have expected the first signals of SUSY already at
LEP2, based on naturality arguments applied to the most minimal models (for example, those with gaugino universality at
asymptotic scales). The absence of signals has stimulated the development of new ideas like those of extra dimensions
and of "little Higgs" models. These ideas are very interesting and provide an important reference for the preparation of LHC
experiments. Models along these new ideas are not so completely formulated and studied as for SUSY and no well defined and
realistic baseline has sofar emerged. But it is well possible that they might represent at least a part of the truth and it
is very important to continue the exploration of new ways beyond the SM. New input from experiment is badly needed, so we all look forward to the start of the LHC.

The most frequently asked questions are: is it possible that the LHC does not find the Higgs particle?
Yes, it is possible, but then it must find something else. Is it possible that the LHC finds the Higgs particle but no
other new physics (pure and simple SM)? Yes, it is technically possible but it is very unnatural.
Is it possible that the LHC finds neither the Higgs nor new physics?
No, it is ''approximately impossible'': that is it is impossible to the extent that the LHC energy and integrated luminosity are considered sufficient for a thorough exploration of the electroweak scale.

\section{Acknowledgments}

As the last speaker, on behalf of all participants,
I thank the Organizers who have done really a great job, in particular Joachim Mnich.
This Symposium presented a very complete picture of our field
in a most confortable setting in the exciting background of the city of Hamburg. 


\begin{footnotesize}

\end{footnotesize}


\end{document}